# Online Labour Index 2020:
# New ways to measure the world's remote freelancing market

Fabian Stephany ✆✦, Otto Kässi✤, Uma Rani❖, Vili Lehdonvirta✦

✦ Oxford Internet Institute, University of Oxford, UK
✤ ETLA, Research Institute of the Finnish Economy, Helsinki, Finland
❖ International Labour Office, Geneva, Switzerland

✆ fabian.stephany@oii.ox.ac.uk

## Abstract

The Online Labour Index (OLI) was launched in 2016 to measure the global utilisation of online freelance work at scale. Five years after its creation, the OLI has become a point of reference for scholars and policy experts investigating the online gig economy. As the market for online freelancing work matures, a high volume of data and new analytical tools allow us to revisit half a decade of online freelance monitoring and extend the index's scope to more dimensions of the global online freelancing market. In addition to measuring the utilisation of online labour across countries and occupations by tracking the number of projects and tasks posted on major English-language platforms, the new Online Labour Index 2020 (OLI 2020) also tracks Spanish- and Russian-language platforms, reveals changes over time in the geography of labour supply, and estimates female participation in the online gig economy. The rising popularity of software and tech work and the concentration of freelancers on the Indian subcontinent are examples of the insights that the OLI 2020 provides. The OLI 2020 delivers a more detailed picture of the world of online freelancing via an interactive online visualisation updated daily. It provides easy access to downloadable open data for policymakers, labour market researchers, and the general public (www.onlinelabourobservatory.org).

# 1. Introduction

Since its launch in 2016, the Online Labour Index (OLI) has provided data and visualizations on the online gig economy to researchers, policymakers, and journalists (Kässi & Lehdonvirta, 2018). We initially developed the OLI as a proof-of-concept with a limited scope. In particular, we tracked data only from large English-language platforms. Although large English-language platforms dominate the overall online freelancing market, there are also important regional markets in other languages that our focus missed. We also did not attempt to visualize many of the temporal dynamics of the market, since we had only been accumulating data for a short time, and we missed data on worker demographics, such as gender.

In this paper we present the Online Labour Index 2020 (OLI 2020), an updated version of the OLI developed in collaboration with the International Labour Organization (ILO). The OLI 2020 makes use of both accumulated time series data and newly obtained data on Russian- and Spanish-language freelancing platforms. In addition, newly analysed data allows us to estimate the freelancers' gender breakdown across geographies and occupations.

The OLI 2020 is made available on the Internet via the Online Labour Observatory, a collaborative website hosted by the ILO and the Oxford Internet Institute (OII) at the University of Oxford. The findings described in this research note and other insights about the online freelancing market can be explored on www.onlinelabourobservatory.org.

# 2. Background

The online gig economy—also known as online labour market or online freelancing market—is understood as the market consisting of platform-mediated work that is conducted remotely via the Internet (Horton et al., 2018). Instead of hiring a standard employee or contracting with a conventional outsourcing firm, companies are using online labour platforms to find, hire, supervise, and pay workers on a project, piece-rate, or hourly basis (Tubaro et al., 2020; Vallas & Schor, 2020). Enterprises from small to large are using these platforms to access skills and flexible labour, assisted by specialized consultants and online outsourcing firms. Dozens of platforms have appeared to cater to different types of clients, workers, and projects, ranging from deskilled microtasks to complex technical projects and professional services (Kalleberg & Dunn, 2016; Difallah et al., 2014). Tens of millions of workers are thought to have sought employment through such platforms (Kässi et al., 2021; Kuek et al., 2015).

Over the past few years, the online gig economy has matured somewhat. Most of the major online labour platform companies have consolidated and listed on stock exchanges via initial public offering, including Upwork, Fiverr, and Freelancer. Recent data from selected countries suggests that while platform mediated remote work is still not a large part of the labour market in any country, it now has some local economic significance beyond what is captured by traditional labour force surveys (e.g. Kässi et al. 2021; Bracha & Burke 2021; Pesole et al. 2018; Collins et al. 2019).

One of the defining features of the online gig economy is that it spans beyond national boundaries. Thus, any measurement effort concentrating on a single country will only reveal a part of the online gig economy's economic footprint (Agrawal et al., 2016; Ghani et al., 2014). Moreover, since a large share of work performed on the online platform occurs in the Global South, it is possible that labour market statistics might fail to capture a considerable share of platform mediated work (see e.g. Kinyondo & Pelizzo, 2018; and Kinyondo; et al., 2019). Motivated in part by this capture problem, the OLI was developed in 2016 as the first economic indicator that provides an online gig economy equivalent to some conventional labour market statistics. Over the last five years, it has been measuring the supply and demand of online freelance labour across countries and occupations by tracking the number of projects and tasks across platforms in real time.

The time series dimension of the OLI allows us to examine the temporal dynamics of online freelance work at a glance. In the last half decade (2016-2021), demand for online freelance work, measured by the OLI, has increased significantly. In early 2021, roughly 90% more projects are demanded via online freelance platforms than in mid 2016 when the OLI started (Figure 1). This equals an annual growth rate of ten percent, which is significantly higher than changes of national (on-site) labour markets, which have plummeted in many countries as a result of the Covid-19 pandemic.

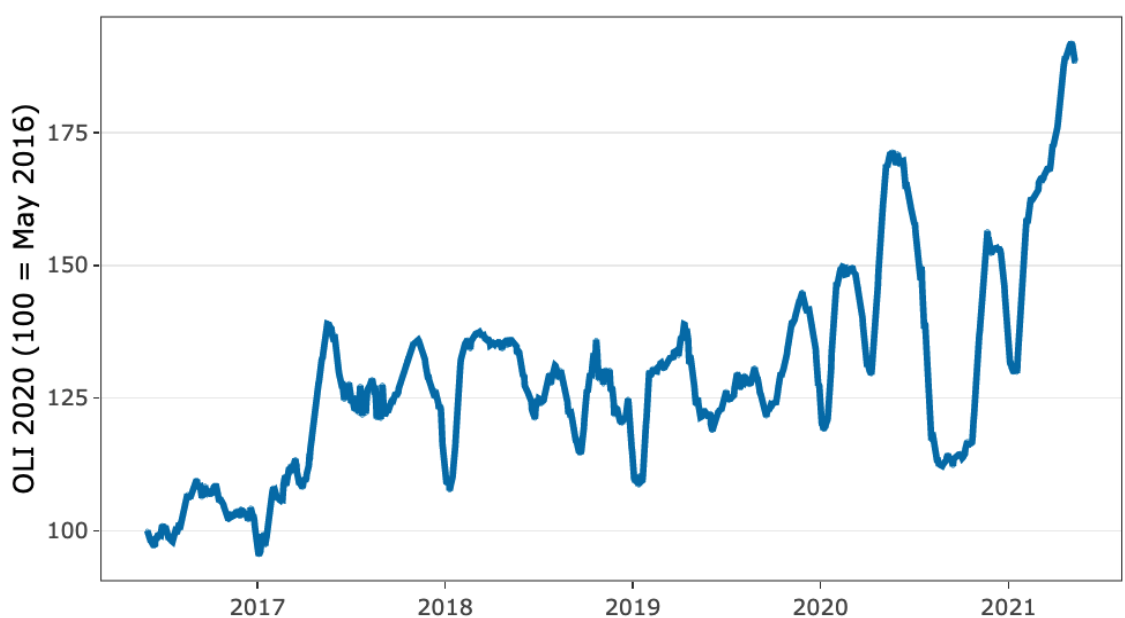

*Figure 1*: Between 2016 and 2021, the OLI registered an overall growth of 90% in online projects while revealing significant seasonal variability over the last five years.

While there is a clear trend increase in the number of vacancies being posted, there is observable fluctuation. Observable dips are visible around Christmas, New Year and summer time, which correspond to seasonal dips in business activity in those sectors most heavily using online services. Similarly, the OLI time series is characterised by pronounced growth spurts in spring (April - June) and autumn (September - November). These may reflect marketing investments made by major platforms, and therefore are probably one-off events. The most pronounced fluctuations on the OLI were tracked in 2020, as a reaction to the global COVID-19 pandemic (Stephany et al., 2020). After an unprecedented drop in demand in March 2020, project requests then spiked to an all-time high in May, followed by a sharp decline in the summer.

## 3. New features of the Online Labour Index 2020

As platform-mediated short-term employment is becoming more popular, online freelance markets have become regionally fragmented. A relevant share of the global online freelance work is mediated via platforms operating on languages other than English (Aleksynska et al. 2021; Shevchuk & Strebkov, 2012). Therefore, in addition to the existing features of the OLI, the new OLI 2020 examines Spanish- and Russian-language platform markets. The OLI 2020 also makes use of OLI's accumulated time series data to make it possible for users to examine temporal shifts in online labour demand and supply, such as the continuously rising popularity of technology work. Similarly, novel data analytics present a snapshot estimate of the participation of female freelancers in the market. Data customisation is an additional feature of the OLI 2020, allowing users to download their desired subset of the data in an easily manipulable form for further analysis and visualization in e.g. Excel.

Russian- and Spanish-language platforms

With the economic maturation of online freelance platforms, regional markets emerged, driven by language barriers and buyers' preference to satisfy their demand for labour domestically (Borchert et al., 2018). Two of the largest regional submarkets shaped by language and domestic demand, outside of the global, English-language domain, are Spanish- and Russian-language freelance markets. To represent the growing popularity of these regionally operating online freelance markets, six new platforms[1] have been added[2]: three from the Spanish-speaking world—freelancer.es, twago.es, and workana.es—and three from the Russian-speaking domain—freelance.ru, freelancehunt.ru, and weblancer.ru.

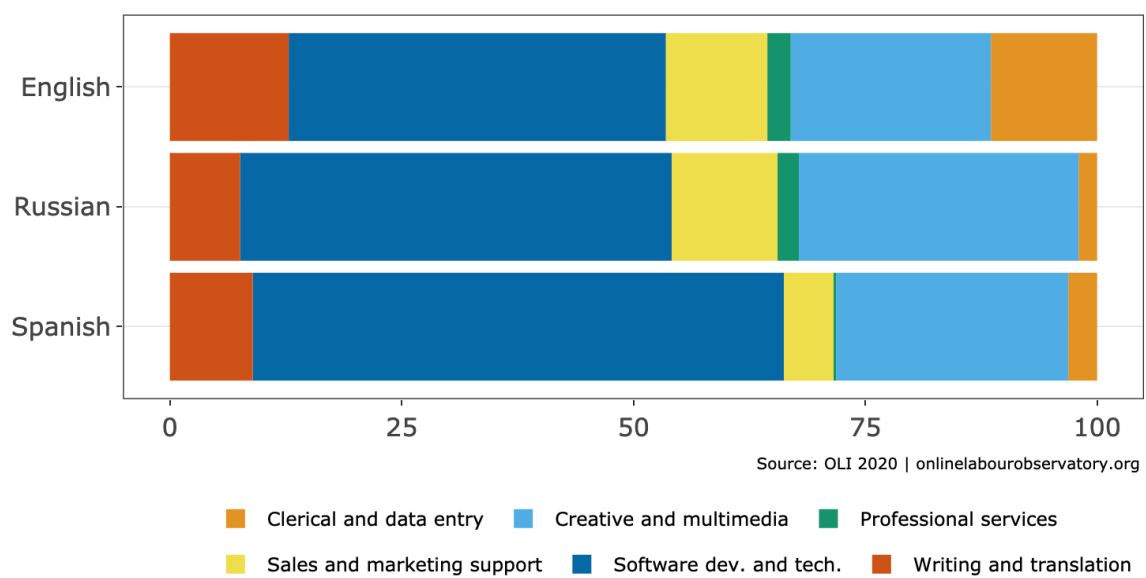

*Figure 2*: *In addition to the original five online freelance platforms, the OLI 2020 tracks the demand of three major Spanish and Russian language platforms.*

As illustrated in Figure 2, the new domain-specific features allow users to explore differences across the three online freelance language markets, such as the significantly larger demand for software and tech work on Spanish language platforms.

---

[1] The initial five platforms monitored by the original OLI are fiverr.com, upwork.com, freelancer.com, peopleperhour.com, and mturk.com.
[2] To maintain the previously acquired time series of the OLI, while avoiding any level shifts in the transition phase from the OLI to the OLI 2020 in September 2020, we employ the economic method of chain linking when adding the new platforms' project count to the existing English-language platforms demand (https://stats.oecd.org/glossary/detail.asp?ID=5605).

## One third of all freelancers come from India

Since the beginning of global online freelancing, countries like India, Bangladesh, and Pakistan have been popular digital outsourcing locations because of their workforces' strong English language and tech skills. The OLI 2020 reveals that over the past five years, the share of workers from the Indian subcontinent has grown rapidly. India's share of the global online worker population monitored by the OLI 2020 has grown from 25% in 2017 to 33% in 2021. For Bangladesh the corresponding numbers are 10% and 15%. This is illustrated in Figure 3, which is produced using OLI 2020's new time series slider.

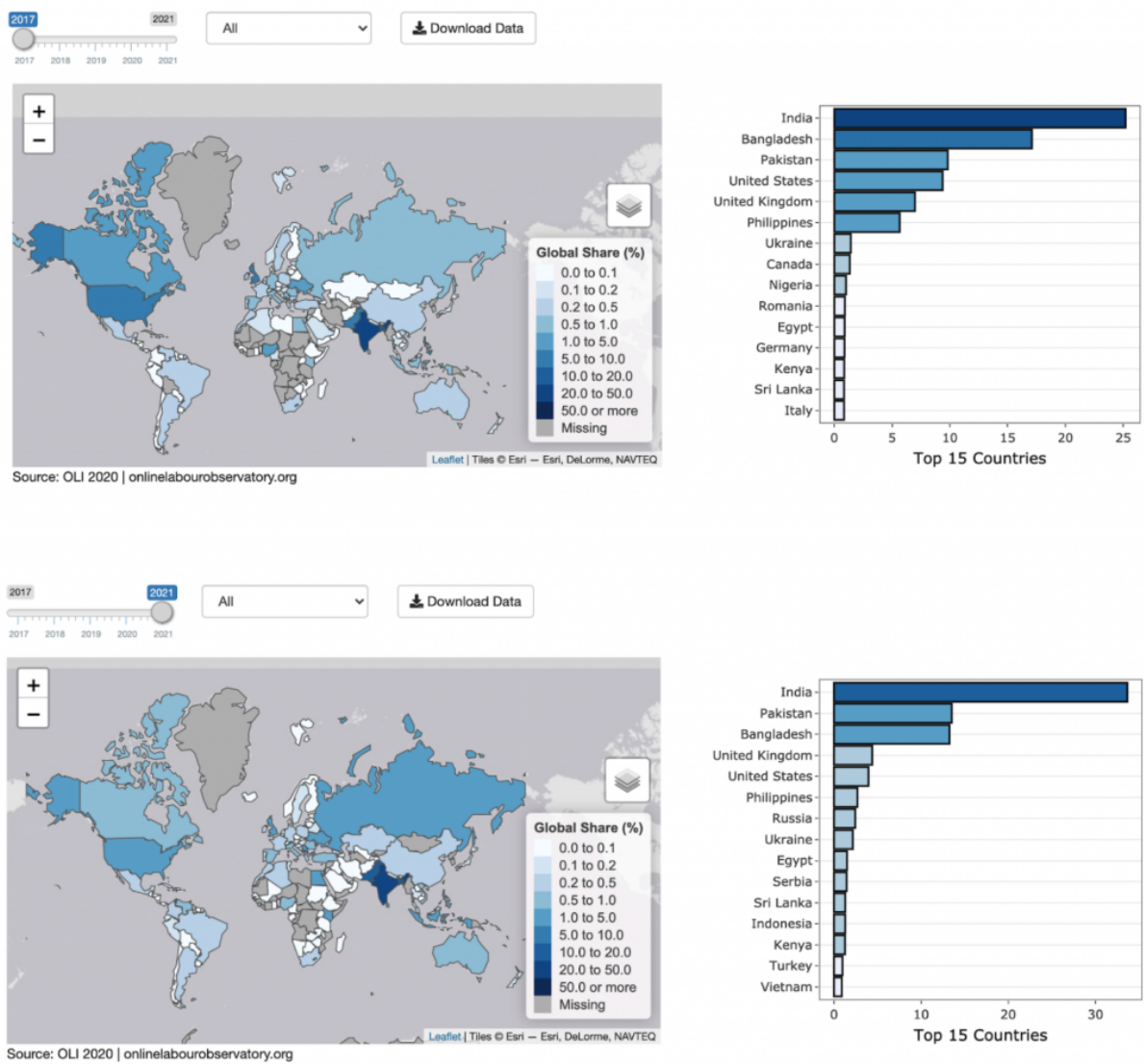

*Figure 3*: Between 2017 and 2021, the share of online labour supply from India has increased significantly. Now, one third of all freelancers are located in India, and more than half comes from India, Pakistan, and Bangladesh.

## Remote software and tech gigs increase

In the last five years, a striking feature of the geography of online labour utilisation is that the occupational demand profiles of the leading countries where the clients are based are rather similar. Clients from all leading buyer countries post most vacancies in the software development and technology category, followed by creative and multimedia, and so on. This is surprising, because the sectoral and industry structures of these countries are very different, as are the occupational profiles of their conventional domestic labour markets. The fact that they nevertheless resemble each other rather much in online labour demand profiles suggests that the demand largely comes from the same industry within each country: information technology, broadly defined. If industries and sectors start making use of online labour in greater quantities, the OLI should begin to show occupational demand profiles diverging in countries' where the clients are located.

The new OLI 2020 indicates that this process might have started. Of the six occupational domains categorised by the OLI, technology and software development has been the most popular group globally and across almost all countries. Over the last five years, however, we observe that the share of projects in this domain has been growing even further. In 2021, 43% percent of all project demand stems from the domain of software development and technology. Comparing 2017 and 2021, software development and technology has become the most popular freelancing occupation in a number of countries, as shown in Figure 4. The new OLI 2020 data also indicates that in some of the developing countries there is also a domestic demand for these activities, which is an interesting trend.

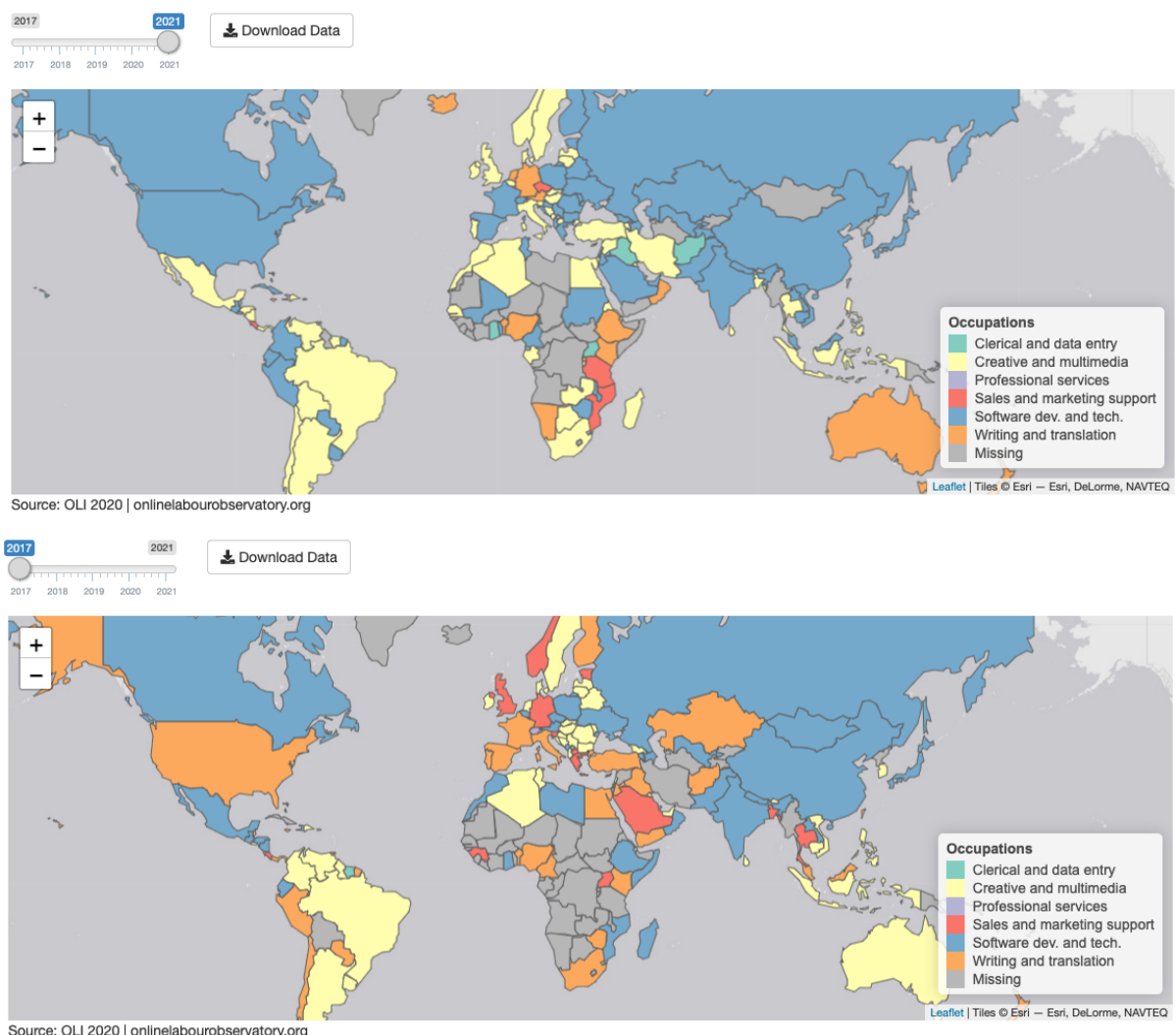

*Figure 4*: Initially, in 2017, the map of popular freelance occupations was more diverse. Now, in 2021, software development and tech work has become the most popular domain in many countries.

Female participation in the online labour market

A new feature of the Online Labour Index 2020 is that the index now provides data on the estimated gender breakdown of online labour supply. The gender breakdown estimates are based on algorithms that make it possible to infer the statistically likely gender of a person from their given name and country of origin (Blevins and Mullen, 2015). The given names and countries are drawn from a subset of the tracked platforms[3]. This is not an exact method, but it provides a view into a dimension of the market that until now has rarely been examined.

Recently, the role of female gig workers is subject of online labour studies, e.g., with regard to gender stereotypes, differences in pay, and occupational participation based on surveys conducted on labour platforms (see, e.g., ILO, 2021; Galperin, 2019; Hunt and Samman, 2019; Liang et al. 2018; Hannák et al. (2017). Overall, our data suggests that female workers make up approximately 39 percent of labour supply in the online gig economy. But there are big country differences. The greatest female participation is in the United States, where women make up 41 percent of the independent online workforce. In India, the corresponding number is only 28 percent. This is especially significant given that India is the largest supplier of online labour overall.

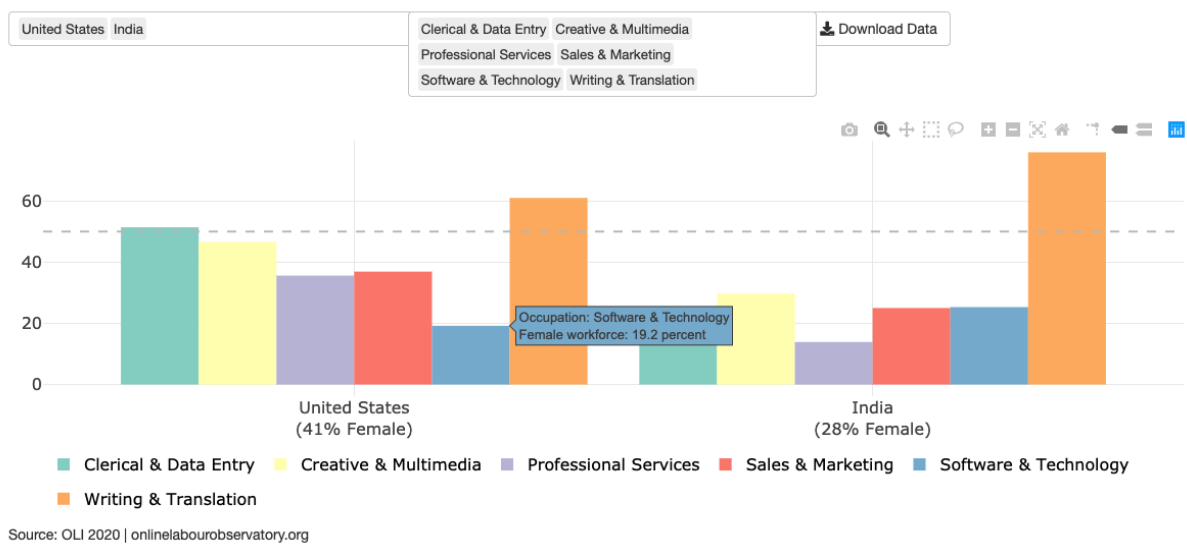

*Figure 5*: *In the United States, a market with 41% female freelancers, more than half of all freelancers in writing and translation work are women but less than 20% of all tech workers are female.*

But when we break down the U.S. and Indian online labour supply by occupational category (Figure 5), nuances in female participation emerge. Although overall female participation is lower in India than in the United States, the order is reversed in the software development and technology category. Only one in five remote tech contractors in the U.S. is female. In India, a quarter of remote tech contractors are female. Participation in tech freelancing thus seems more equal in India than in the United States. Still, in both countries, the largest proportion of women by far is found in writing and translation work, suggesting that online occupations remain somewhat segregated by gender.

---

[3] We estimated the gender for a subsample of 14,838 freelancer profiles drawn from upwork.com and peopleperhour.com, retrieved between October 2020 and January 2021.

# 4. Conclusions and future development

There is an ongoing discussion in the policy arena with regard to the number of workers who access platforms for work and whether this is a significant trend and a new form of employment opportunity. As platforms do not publish the number of workers who access their platform for performing tasks on a regular basis, it is difficult to have a reliable estimate. The Online Labour Observatory addresses this issue partly by showing that on some of the major digital labour platforms, the demand and supply of labour has been steadily increasing since 2016, and that platform work cannot be ignored and is an emerging phenomenon in the labour markets. This data is relevant to policy makers as it provides an overview of the people who are dependent on these platforms by gender, country and region, and also by occupations. It also shows that platforms operate across multiple jurisdictions as clients from different parts of the world post tasks or projects on these platforms, which are completed by a global labour force.

As the working conditions on these platforms are largely regulated by terms of service agreements, they tend to characterize the contractual relationship between the platform and the platform worker as other than employment (ILO, 2021; Wood & Lehdonvirta, 2021). As a result, platform workers cannot access many of the workplace protections and entitlements that apply to employees raising concerns. The growing number of workers accessing work on these platforms suggests that governments need to address some of the challenges arising from platform work, including that of the ambiguous employment relationship. This would require developing policies to ensure that workers receive universal minimum labour and social benefits and rights, and to find regulatory solutions as these platforms operate across multiple jurisdictions through international policy dialogue and coordination.

A potential future direction for the development of data collection efforts in this area, including the OLI 2020, is a focus on greater geographic granularity. Braesemann, Lehdonvirta, and Kässi (2020) use subnational online labour market data to show that relative to population, freelancers in the rural counties of the United States appear to be doing more work online, and also doing work that is of a higher skill level than freelancers in urban counties. Future extensions of the OLI 2020 could make subnational units, rather than countries, the smallest geographical unit of analysis, providing an even more detailed representation of the spatial dynamics of the world's remote freelancing market.